# Balancing Innovation and Sustainability: Learn the Potential Impact on the Environment of Bitcoin Mining

**Mohammad Ikbal Hossain**
**Emporia State University**

**Muhammad Imam Hussain**
**Binghamton University**

**Ayesha Arobee**
**Emporia State University**

**Md Nasim Fardous Zim**
**Emporia State University**

**Md Bahauddin Badhon**
**Emporia State University**

**Shahanaz Akter**
**Southern Utah University**

*Convergence of technological development and sustainable development principles in the context of Bitcoin mining, are the cross-cutting issues this study is poised to address. Given the increased usage of Bitcoin, there have been increasing environmental concerns about electricity consumption and the carbon footprint of the mining processes. The study emphasizes the primary Bitcoin mining steps about how the process becomes energy-ravenous due to the proof-of-work mechanism attached to it. It also evaluates the potential environmental impact arising from mining activities with a primary focus on emissions and e-waste. Economic models demonstrate that the energy level that is used in Bitcoin mining is equivalent to the energy used by some countries, thus raising the need for appropriate management as such levels of energy use might be unsustainable. This paper also aims to predict modifiable changes in reducing potential environmental problems related to bitcoin mining by supplying some energy conservatism alternatives: energy-efficient mining equipment along with the utilization of solar energy, wind energy and water energy. Moreover, the report presents the current activities and their successes in encouraging sustainability practices in capital consumption or minimization of waste materials, policies for carbon neutrality and a framework for efficient e-waste management. There are legislative, institutional and other control measures along with incentives. They are aimed at reducing the environmental degradation caused by Bitcoin mining. The paper argues that it is possible to develop policies or employ technical solutions that would mitigate the environmental consequences of bitcoin mining without compromising the technological advancement of bitcoin mining.*

## INTRODUCTION

In management and business, innovation and sustainability are two terms repeated most frequently. The recent phenomena of such as the rise of crypto currencies have shown that there is need to discuss the junction between innovation and sustainability in context of bitcoin mining for instance it is necessary to strike a balance between environmental sustainability as well as technological innovation (Kayani & Hasan, 2024; Gunay et al., 2023). As demand for cryptocurrencies continues to rise, so does energy consumption in its mining operations (Islam et al., 2022). In 2009 Bitcoin was founded as the first digital currency; however, after remaining skeptical for some time it has been increasingly accepted into mainstream economic activities, including investments, finance, real estate and luxury industries. These cryptocurrencies—Bitcoin or Altcoin for example—arise from a competitive process called mining where their block chains get verified (Spurr & Ausloos, 2021). Among other things, it causes sustainability concerns such as theft and illegality coupled with heavy energy use or depletion of minerals therein. Remarkably, much is about environmental pollution, as we recognize that there have not been enough sustainable practices that can safeguard our ecosystems from collapse.

Bitcoin's most pointed out impact is through its various exchanges' ecological footprint (Sarkodie et al., 2022). This footprint results from an accumulation of transactions frequency and computation cost but not induced by price or metal recycling in computers. Bitcoin's ecological footprint is greater the larger the blockchain sizes (Bajra et al., 2024). However, skillful implementation of Blockchain technologies may lead to increased digitalization and economization of energy. The latest use of blockchain technology is based on combining renewable energy sources or creating energy communities that will support private decentralized models with mixed control. This interaction between the ecological and digital economics requires a balancing point between the necessity for new tool or virtuous applications innovation and the ecological footprint of their products or services generated by such innovative actions, which had been completely omitted until now. This work aims at developing on how to follow critically such trajectories worldwide. Consequently, it is urgent to confront these two arguments by reflecting through this angle to avoid adverse consequences that have been previously warned of and documented.

## UNDERSTANDING BITCOIN MINING

Bitcoin was introduced as a digital currency in 2008, and it uses blockchain technology which verifies and authenticates transactions using decentralized ledger systems (Arora and Nagpal, 2022). Unlike the traditional systems where central authorities (governments and banks) controlled transactions and fees, this technology relies on the combined power of many computers across national borders with contributions from peers around us all. This means that data can be sent and stored without the need for central authority oversight. It is maintained anonymous and in a decentralized manner by miners contributing computing resources for validating transactions. The open-source Bitcoin protocols' design helps facilitate the heavy encryption and its lack of centralization (Ghosh et al., 2020). Newer Bitcoin protocols serve to maintain blockchain block time stability and guarantee that only legitimate transactions are added to a block before incorporating fresh requests in an accepted decentralized way.

Miners in the decentralized blockchain infrastructure are those that adore books whose purpose is to secure the network ensuring that any new blocks contain only valid transactions in them. Many transactions directed toward the Bitcoin network are concurrently processed, but permission checks allow only a small batch per ten minutes (Erdin et al., 2021). These new block requests must then wait for the previous ones to be verified although one miner can append additional requests by successfully resolving a proof-of-work challenge, which leaves other miners with the responsibility of solving the subsequent round of similar puzzles (Maleš et al., 2023). To guarantee this, every new block should include the subsequent hash of previous new requests so that it displays accurate information, which involves all relevant transactions. At times one may refer to them as mining pools which are basically a few miners.

## Overview of Bitcoin and Blockchain Technology

Bitcoin is a decentralized digital currency and a payment system. It operates on the principle of peer-to-peer and has done away with central authorities and middlemen. The brilliant technological breakthrough underlying Bitcoin is blockchain technology. Blockchain technology is a series of networked computer programs or software that store data about a given digital currency's transactions anonymously and transparently simultaneously (Toorajipour et al., 2022). All transactions must be secured in a way that once done, they cannot be undone, and everyone can agree on the status of the book at any given time. An analogy can be made of each block in a blockchain being a page of a history book, with all blocks having to be linked to each other using a special string called a hash (Lyu et al., 2020).

One important aspect of blockchain technology is its unique properties, making it possible to be used in many other areas beyond digital currency alone. To start with, it is immutable to mean that once a digital ledger has been updated, it is almost impossible to change anything in it (Madaan et al., 2020). Moreover, it is transparent because every member of the network can view all transactions before they are made public on the network. Each transaction is recorded in the digital ledger under the “pseudo-anonymous” condition, i.e., with no names or SSNs but wallet addresses, which can be traced to individuals with complete access (Azman and Sharma, 2020). This way, the wallet address becomes crucial in connecting network users to their identities. When processed into the blockchain, transactions can never be modified, disappear, or repeated. Thereby each node stores the ledger itself enabling robustness of the network against failure by distributing it all over the nodes. Block creation plays an important role in the blockchain network and helps solve the double spending problem.

## The Process of Bitcoin Mining

Mining aims to generate new bitcoins and validate transactions made by users. E.g., if you buy coffee with bitcoins, the network must verify that your account contains enough money to settle this transaction (Tsang and Yang, 2021). In addition, one of the coffee drinkers will send some bitcoins to the child sex offender in Montreal. To achieve this, all network participants must receive a transaction request from the user and then one of them should solve a mathematical problem (Huang et al., 2021). The puzzle solution is freely available on the network, so anyone can verify if the block has been worked out correctly. Each such commitment is bundled with all the transactions into a block, treated as a long broadcast list which is connected through some kind of cryptographic fingerprint with the one that was before it This way we can be sure that all transactions submitted to the network have been included in it. Bitcoin’s energy consumption over time can be spotted in Figure 1; this will form the basis for further discussions on its environmental impact. Miners do not solve puzzles all at once; instead, they solve them in an iterative manner. They systematically change puzzle input bits and check if the output matches a target exit value (Chaurasia et al., 2021). This approach makes random guesses after random guesses with each one taking a split second for a newcomer, requiring a minimum of 10 minutes to make one guess (Cao et al., 2022). This mechanism is called proof-of-work: it is essentially a difficulty setting for the block puzzle. Here, the proof-of-work is like buying difficulty settings in computer games. In fact, proof-of-work ensures that one single block cannot be spammed; hence, the blocks have to be spaced out to 10 minutes apart.

**FIGURE 1**
**BITCOIN ENERGY CONSUMPTION OVER TIME**

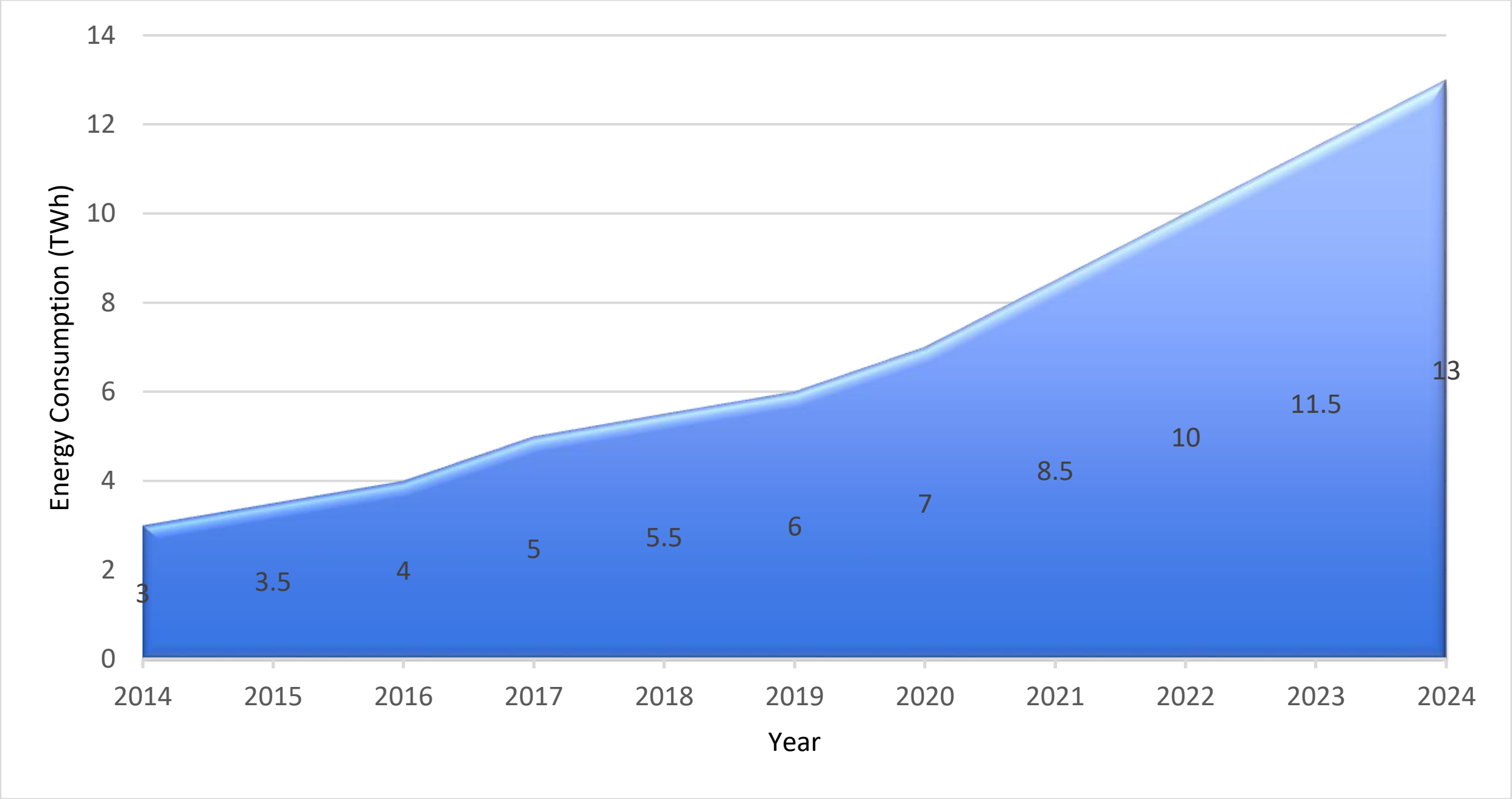

*Sources: Cambridge Bitcoin Electricity Consumption Index (CBECI)*

With increased hardware power, mining hardware can generate a higher hashrate quickly. You should note that, the reward for a block (besides the block reward) is more worth when the block is at full power (Alharby, 2023). Mining bitcoin requires solving hard mathematical problems to verify transactions and publish them on the public ledger.

## ENVIRONMENTAL IMPACT OF BITCOIN MINING

Bitcoin mining's environmental impact has become a matter of concern in academic research and media reports alike. The major concern is the high amount of energy it consumes considering the estimated global energy used during mining activities (Gallersdörfer et al., 2020). Also, many Bitcoin mining centers are situated in places with high carbon intensities with concerns arising from their high energy consumption over the revival of fossil fuels (Schinckus, 2021; Donovan, 2023; Finney, 2024). The discussion connects socially embedded environmental issues concerning the availability as well as accessibility of energy resources in a given location and the impacts on the local or regional environment through high energy use. The enormous carbon footprint resulting from this process is often seen as an undesirable consequence of a sector that faces increased scrutiny regarding global matters and the relevance of cryptocurrencies in the global economy (Pagnuelo, 2019). Moreover, e-waste generation annually via bitcoin mining emanates from the production of rapidly outdated hardware utilized during this operation (Jana et al.2022; Jana et al.2021). Consequently, an increasing number of scholars are investigating how harmful it is to our environment, demonstrated by a sharp increase in the total number of academic works on such topic over the last years. This research takes several standpoints on these impacts' exact magnitude (De Mello et al., 2018). We should also be aware that many researchers have argued for improved models on how bitcoin impacts the environment given that there has been rapid growth in transaction volume leading to transaction verification. Such a model should be built upon more accurate data thus making it possible to completely comprehend how our environments are affected by mining activities within a locality. This can be done by establishing better future potential effects through an understanding of their current timeline in addition to dealing with possible circumstances that may arise later (Awan et al., 2021; Mendoza et al., 2020; Varseveanu et al., 2020).

Two major factors are responsible for bitcoin mining's environmental impact. One of them is the exorbitant electricity cost carried by miners or the validating nodes. They do this by employing a number of machines to tackle complicated algorithms and utilizing old hardware that quickly loses usability due to its complex exploitation (Badea & Mungiu-Pupăzan, 2021; Siddique et al., 2023). This consumes enormous amounts of energy and gives rise to enormous amounts of greenhouse gas emissions which alter earth's climate. The other reason is because this process creates so much electronic waste i.e., outdated hardware that regular hardware facilities cannot use for mining cryptocurrencies like bitcoin (Heinonen et al., 2022; Gola & Sedlmeir, 2022; Alfieri et al., 2023; Goel et al., 2024). In order to earn cryptocurrencies, it is now necessary to have newer and more powerful computers which mean more hardware trash that comes out as a result of Purchasing new computers each time the algorithms need re-solving. The other issue with using physical facilities is the need for space, cooling systems, and electricity further raising the cost for Bitcoins mining operations.

**Energy Consumption and Carbon Footprint**

Several computational and theoretical models estimate energy consumption of Bitcoin mining. However, no consensus has been reached in literature on this issue yet. The electricity consumption in the Bitcoin network per year is equivalent to nations like Colombia, Argentina, Finland or Switzerland (Náñez et al., 2021; Corbet & Yarovaya, 2020; HAROONI, 2023; Okorie et al., 2024). The carbon footprint of Bitcoin transactions depends majorly on the carbon intensity of its energy. Available data show that at least 39% of global electricity used by bitcoin miners comes from renewable sources (Malfuzi et al., 2020; Bastian-Pinto et al., 2021; Niaz et al., 2022; Hallinan et al., 2023). Figure 2 shows visually where the mining activities cause more emissions in different countries. Moreover, most of this renewable power relies on nonrenewable sources like hydro or biomass or even nuclear power, which raises questions about their adverse social and environmental effects (Voumik et al., 2023). Sometimes, you would find bitcoin mining happening in areas with plenty of clean energy but very little industrial demand for electricity; thus, enabling these countries to handle its intermittency during off-season (Bukhari et al., 2024). However, these conditions are hard to come by. There are different types of power plants with different levels of carbon intensity and for bitcoin mines they use the cheapest one locally. Renewable energy is usually used because it is cheap (Мельник et al., 2020; Razmjoo et al., 2021; Luderer et al., 2022; Bogdanov et al., 2021). This kind of "spill-over" increases the carbon intensity for other end users or exports. There are some carbon-intensive plants that are only operating because they are idle when they are still operational. Moreover, it is difficult to scale renewable energy sources due to weather dependability; coal, natural gas as well as nuclear being major primary energy sources that have several demerits in terms of greenhouse gas emissions together with waste products (Wang et al., 2023).

**FIGURE 2**
**CARBON EMISSIONS FROM BITCOIN MINING BY COUNTRY**

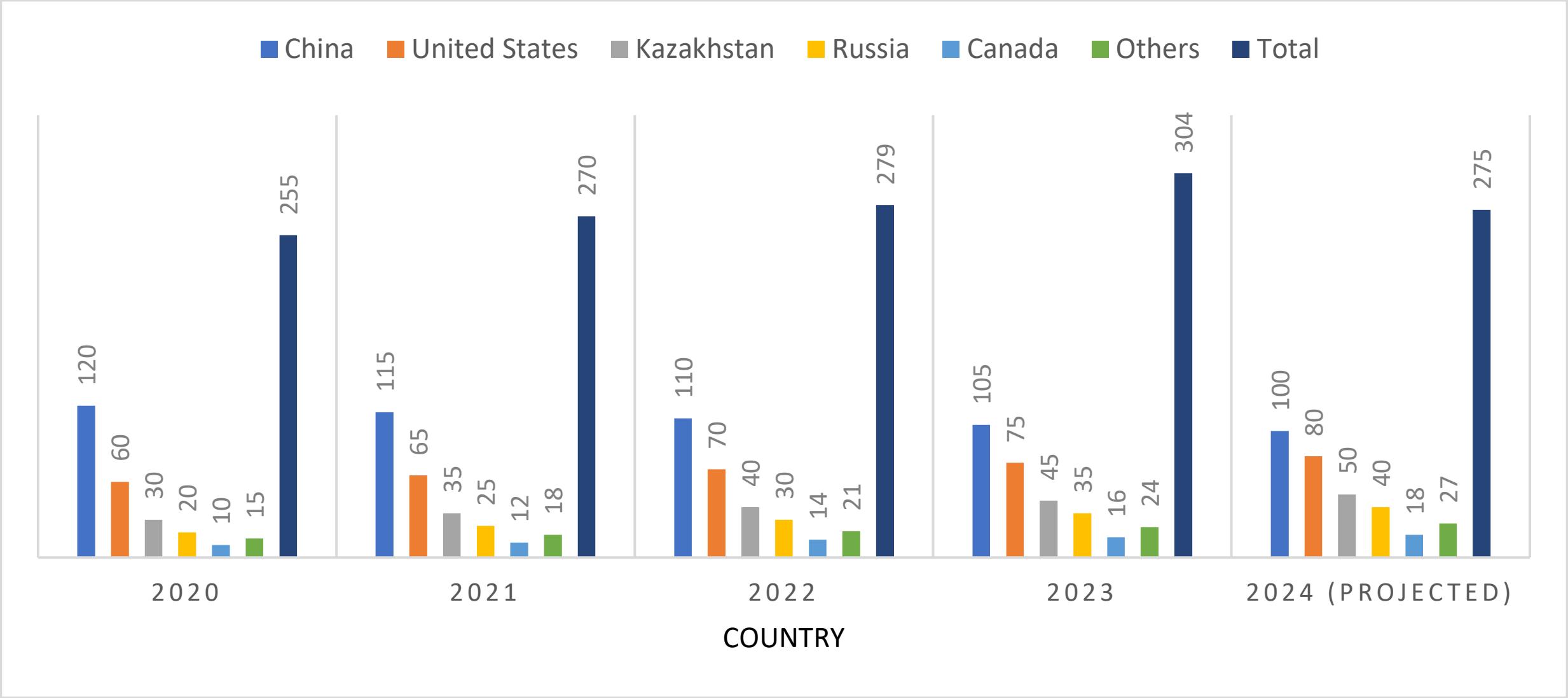

*Source:* ***China Data (2020-2023):*** *CBECI Reports;* ***United States & Others:*** *Industry reports and IEA estimates*

**Generation of E-Waste**

Bitcoin mining hardware has an average lifespan of 1.29 years. Therefore, after the mining process is over, there is a need to recycle or dispose of the hardware used to mine Bitcoin (Ferdous et al., 2021; Sandberg and Chamberlin, 2023). Some individuals who engage in mining will pull out certain parts in their old hardware such as large fans for use in other places. Because it's difficult for miners at large to get rid of their obsolete hardware well, they end up causing electronic wastes (Ghulam & Abushammala, 2023). For instance, if one ASIC machine was properly disposed off by its miner annually, then we would expect it to produce a litter ranging from 10-90 kg of weight per ASIC machine when weighed against a lifespan within 18—32 years (Bachér et al., 2022; Farjana et al.2023; McNally and Kolivand, 2024). While these figures are only speculative, it is apparent that more profound e-waste can be observed over all phases of its life, from the ASIC machine's production to its disposal after becoming obsolete.

Older technology becomes obsolete, newer technology becomes faster. As a result of this upgrade cycle, older technology often becomes outdated as it is no longer able to make profits and hence out of date. This defunct hardware is then neglected. For instance, after Liu et al. (2020), He et al. (2021), Peserico et al. (2022), Martins and Gresse Von Wangenheim (2023), we got many 1,448 old miners with 28.9 MW of total energy usage. This means that the coins they used to mine have become outdated, contributing to future e-waste directly. The infamous 1448 obsolete machines constitute over 55% of all e-waste recorded despite constituting 12% of hardware pieces operational at any given time. Consequently, in terms of physical volume consumed for storage purposes by these machines, as they age, the amount of waste in the form of e-waste continues to rise. Although people mostly keep them in large warehouses or immersed in oil, there are some who may opt for it depending on circumstances. Compared with money generated through abstract mining activity from all functioning CPU, it has an approximate value of about 5.1 million dollars without any record. However, little literature specifically addresses e-waste occurrences concerning WEEE within the mining sector though research on forecasting e-waste from general waste electronic and electrical (WEEE) is ongoing. Internet community of late has started writing some online advisement on mining industry electronic waste recycling however with time these activities seem to have slowed down too much (Erdiaw-Kwasie et al., 2024). From Figure 3, e-wastes have been growing exponentially with time, a fact that underscores the magnitude of challenge posed by hardware turnover.

**FIGURE 3**
**BITCOIN MINING E-WASTE GENERATION**

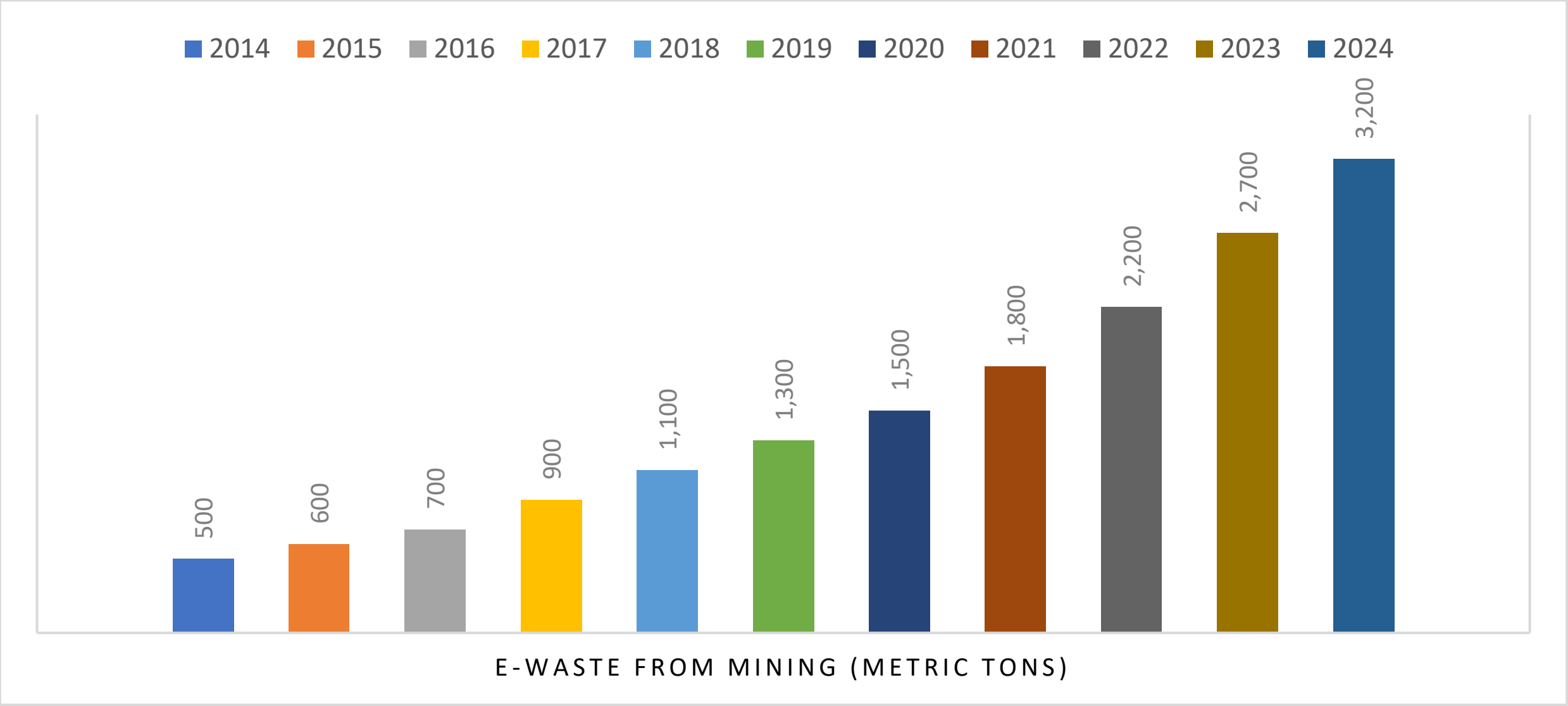

*Source:* ***Jana et al. (2021, 2022;*** *Industry reports and IEA estimates*

As the popularity of Bitcoins increases, it is expected that e-waste generated by mining hardware will also rise in exponential terms alongside the power consumed. Therefore, to safeguard resilience, the focus should not only be on benefits versus costs analysis but also on creating alternative mining approaches considering its growth rate (Chen & Ogunseitan, 2021). We propose that attending to the disposed equipment together with the e-waste emanating from it is closely related to addressing the increased energy usage by cryptocurrency-specific microchips (ASIC machines). For instance, 1,448 defunct miners – representing 83% of all working machines—awaited retirement within its premises over a passive period of 1.29 years. Consequently, it would be against our best judgment to use electricity hitherto earmarked for any part or whole of the fossilized devices since their operation produces much lower digital currency densities for each square meter occupied in such than crypto warehouse pays for power. Stockpiling idling machines, therefore, appears as an avenue for possible future monetization while appreciating that energy investments with zero currency yield are currently costly (Suárez, 2023). Operational plus defunct hardware means that a loss of power amounting to 20 GWh or 69 GWh after 1.29 years is incurred from the day it was opened (Cooke & Xu, 2024).

## CURRENT SOLUTIONS AND INITIATIVES

Only a few publications present sustainable practices that can reduce the environmental impact and thus future projections. Legalization of marijuana may have some unintended environmental consequences according to one such study of greenhouses (Klassen and Anthony 2022). Several Californian growers were using a powerful pesticide despite its ban since it is harmful to birds. As of 2009, no regulation or oversight monitored energy use (McClean & Pedersen, 2023). However, if one thing remains certain about them all, it is that they require clean mining that does not compromise any ecological principle at the expense of today's gains (Weber & Koomey, 2022). The only way to bring down emissions is through using renewable energy sources; anybody who desires to see this happen would have no choice other than to integrate these two technologies within their bitcoin mining operations (Albaron, 2020). At the practice level, there are several options available to major miners; these include putting at least some percentage of their production on green energy and switching to more energy-efficient hardware (Neus & Boack, 2022). This can be attributed to the fact that unless otherwise stated, most of them are usually tied into the central grid, meaning they exist in geographical locations that take part in trading carbon offsets or similar arrangements (Sharma, 2022). Few Bitcoin miners have followed suit, but very high electricity consumers have put more effort into becoming efficient in their energy consumption than others. The industry leaders have promised to design better Application-Specific Integrated Circuits (ASICs) that consume less power while doing calculations (Milutinović et al., 2021). This can lead to about two times hash rate on mid-term horizon under conditions in which computational capacity does not double but energy consumption rises by 50% approximately (Desislavov et al., 2021). The self-contained ASIC is 40% more power-efficient when compared to its nearest competitor (Suleyman, 2023). This gap can be filled by better cooling technology so that at maximum loads the inside temperature should not exceed 50 degrees Celsius. Such systems are conveniently air-cooled for these purposes without requiring entire room immersion cooling systems as before. Businesses are striving for data mining center ambiances that are identical to those of banks and negotiating with ecological societies (GURGU et al., 2021). This might reflect attempts to respond to a growing body of evidence documenting negative environmental effects or serve as a new wave of innovation aimed at promoting green mining technologies. Discussions are in progress about the development of mining best practices and industry-imposed standards for bitcoin. These initiatives are likely to attract responsible business behavior since these companies' impact may prove critical in the future.

**Integration of Renewables**

To produce environmentally friendly and cost-effective electricity for Bitcoin miners in North America and worldwide, many of them have embarked on an integration process involving solar, wind, hydro and other renewable resources. Some of these miners have pioneered the adoption of solar power in the context of large-scale mining operations (Pouresmaieli et al., 2023). Solar energy use by large-scale mining sector is mostly through on-site or offsite solar deals, which are made possible by long-term alliances with nearby utilities. In addition, miners are constructing mining data centers that incorporate renewable sources, including solar technologies (Sovacool et al., 2022). The co-location and crypto-mining facility's owner and manager had plans to attach solar panels and wind turbines on its roofs soon. Besides, clients of distribution utilities have been incorporating grid-tied photovoltaic units into their own cryptocurrency mining activities.

When green power sources are incorporated in cryptocurrency beyond Bitcoin, it can become a great opportunity. Reducing gas emissions and hash carbon output per reduce the greenhouse gases helped so much in making this industry environmentally friendly, but more importantly, this can lead to savings (McNally and Kolivand, 2024). Nonetheless, impediments hinder the promotion and acceleration of a mining baseload that relies on only green energy. Most data center operators remain connected to grids despite some choosing to purchase regional green power certificates, the rest of these centers cannot access Power from RESs (Khosravi et al., 2024). Also, 100% renewables cannot guarantee an uninterrupted supply of electricity 24 hours a day, so those wishing to go off-grid must consider investing in storage solutions (Kuznetsova & Anjos, 2020). An event that such requirement should occur might open little and brief asymmetrical trading opportunities but may present difficult challenges because just part of it could be met using Brian energy sources such as what is available or stored from sunshine in batteries during daytime only. Figure 4 shows some switch (or continuing lack of shift) toward using green energy for mining Bitcoin, serving as an indicator how far (behind) this sector has gone on cleaner production.

There is great potential for increased use of clean energy, including solar and wind power in general, particularly for integrating bitcoin mining among other renewable energy sources. Consequently, this kind of initiative provides an eco-friendly mining involving less emission of gases that cause the greenhouse effect and less carbon dioxide per hash alongside other benefits such as coin saving over time (McNally and Kolivand, 2024). Despite these facts, promotion and acceleration of baseload mining based on renewable resources has its own challenges though. Most data centers are connected to the electricity grid, but some may choose to purchase green energy certificates specific to their local areas only; therefore, not all data centers can access electricity from renewable sources or those that cannot be sustained (Khosravi et al., 2024). Furthermore, contrary to popular belief, renewable integration does not support the operation 24/7; hence those planning to go off grid should explore possibilities of energy storage systems like battery technology development (Kuznetsova & Anjos, 2020). Direct sun energy can work for Bitcoin miners under certain conditions since it is possible to temporarily arbitrate using other sources. However, this is not the case when it comes to non-stop power consumption because most other types of renewable energies especially solar electricity generated during day hours without back-up will not solve the problem of 24/7 electricity for Bitcoin miners. Figure 4 shows either transformation towards the use of renewable energy or not in Bitcoin mining industry thus indicating how much change has occurred.

**FIGURE 4**
**RENEWABLE VS. FOSSIL FUEL ENERGY USAGE**

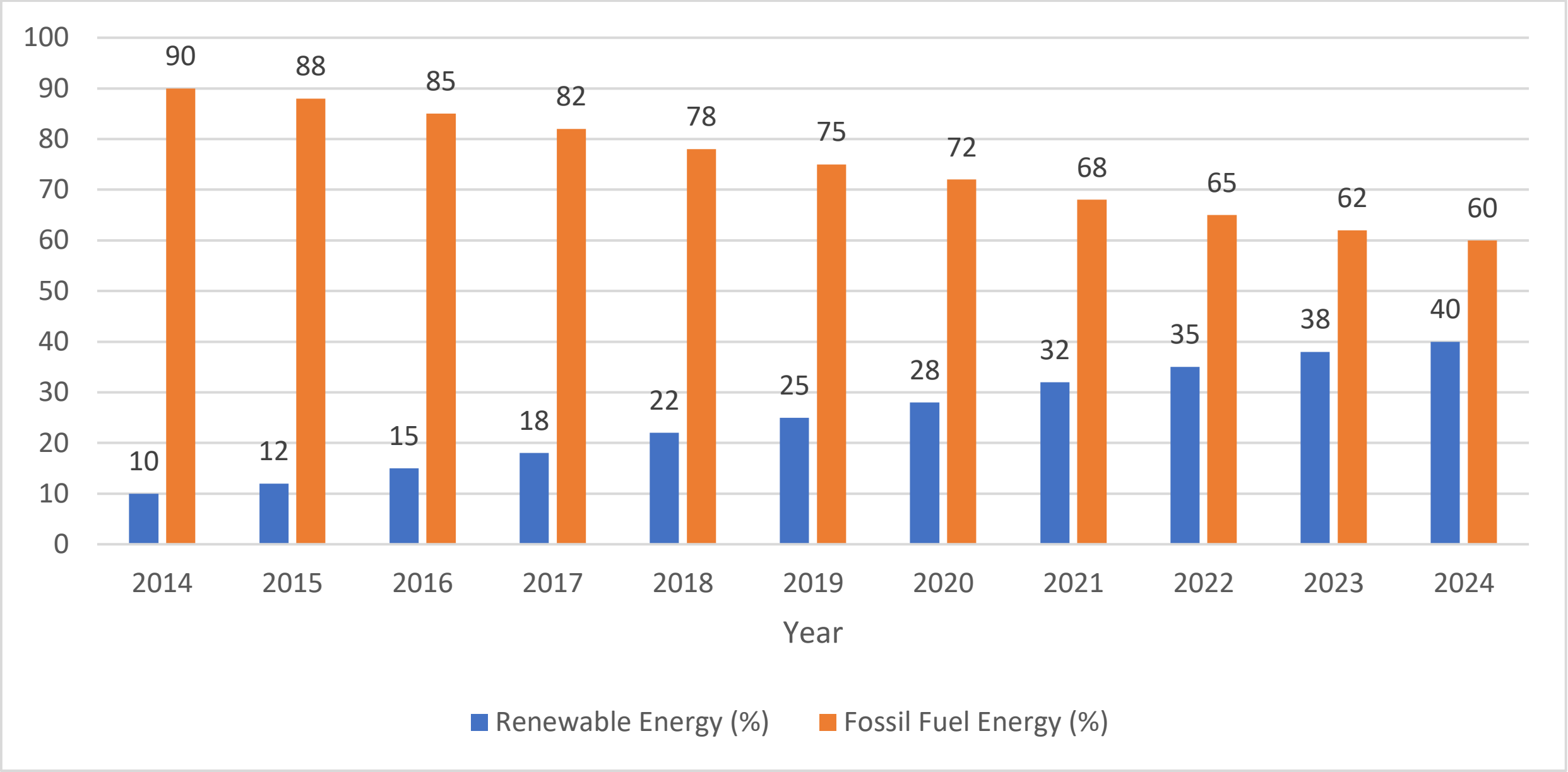

According to Lu et al., (2020) governments can play an important part in creating incentives for renewable energy sources. In the United States, the zero-emission nuclear power and electricity production tax credit program increases to more than $30 per tonne for qualifying electricity producers. Other nations may offer miners possible reimbursement for additional investment or production tax credit as part of their renewable energy investment schemes for private enterprises. In the same vein, the core policy direction of advocating for clean, available, and affordable green power and encouraging the sitting of large energy-generative renewable projects to back larger mining farms exists (Qi et al., 2020). Ontario province in Canada is today a leader in North America in developing mining and energy data center and infrastructure.

**Efficiency Elevations in Mining Hardware**

A significant aspect of technological advancement in the bitcoin network has been the continuous improvement in energy efficiency of mining-specific hardware (Jabłczyńska et al., 2023). The early hardware such as CPUs, GPUs, and FPGAs were quickly replaced by ASICs which as models improved have become more efficient and less wasteful (Fryer & Garcia, 2023). There are even underground ASIC data centers where it is required that ʽall that waste heat has to go somewhere; it can't just be unused.' Recent developments on least energy emitting equipment have been made for sustainable power use. Its origin dates to mining hardware that uses not just micro-processing-based equipment but also uses Breadboard Chips i.e., mining setups employing electronic circuits on a hardware breadboard. Emerging Environmental policy initiatives have also seen an increased interest in developing very energy-efficient cryptocurrencies. For environmentally friendly cryptocurrencies, efficient hardware will be crucial. This includes creating efficient hardware with lesser energy footprints (Yazıcı et al., 2023). For example, while the 2020 model of S19j achieves 7.5 J/GH due to higher productivity, the same metric for the S9 is roughly 34.5 J/GH alike as an average of 21.0 J/GH which we will employ as the operating energy in our paper.

Furthermore, several manufacturers are investing in R&D activities to produce energy-saving equipment including integrated and specialized electronics or microprocessors. It is worth observing that there are financial gains in lean hardware for mining; if fewer miners use electricity still there will be more capital for buying machines (Bikubanya & Radley, 2022). Energy-efficient hardware is a concomitant to Bitcoin electricity network's hidden health properties. In other words, better hardware can produce higher hash rates with identical use of power and investment hence reducing overall carbon footprints (Xiao et al., 2023). However, this body of evidence shows that mining more coins with high computationally difficult puzzles is compatible with zero or low carbon networks. In future this will improve credibility among financial institutions which are reluctant to embrace any digital currencies. Nevertheless, mining technology progresses require qualitative changes (structural review) along with data processing (Douaud et al., 2022).

**TABLE 1**
**COMPARISON OF ENERGY EFFICIENCY IN MINING HARDWARE**

| **ASIC Model** | **Power Consumption (J/GH)** | **Hash Rate (GH/s)** | **Lifecycle (Years)** | **$CO_2$ Emissions per Bitcoin Mined (kg)** |
|---|---|---|---|---|
| S9 | 34.5 | 14 | 2 | 150 |
| S19 | 21.0 | 95 | 3 | 90 |
| S19j | 7.5 | 110 | 3.5 | 50 |
| Antminer Z15 | 5.0 | 170 | 4 | 35 |
| Future Model X | 4.0 | 200 | 5 | 30 |

*(Source: Manufacturer Specifications: Bitmain; Framework for sustainable energy use in mining (Yazıcı et al., 2023))*

Compared to electricity use, hash rate, lifecycle and carbon emissions from some models of ASIC chips gathered to produce Bitcoin, Table 1 is created. Progress of mining hardware from models of past generations such as the S9, characterized by only 34.5 J/GH efficiency rate and 150kg $CO_2$ emissions per mined bitcoin, to current generations such as Future Model X, which could only be 4.0 J/GH efficient, with only 30kg $CO_2$ emission, is displayed here. All these newer models – S19j, Antminer Z15- show improved energy utilization and longer lifespans that are indicators of attempts within the industry aimed at reducing their harm to the environment. This therefore means that these newer models are more power efficient and significantly lower the carbon footprints related with Bitcoin mining, underscoring how technological progress can play an important part in promoting sustainability in the cryptocurrency sector.

Hence, if there is to be a widespread introduction of new “greener” cryptocurrency technology, we recommend policy makers and lawmakers initiate discussions on structural, social, and financial experimentation in addition to the technological and business experimentation. One potential research avenue is the comparison between traditional and hydroshacks. According to our field data, hydroshacks may already deliver lower operational costs in high-energy regions, hence environmentally friendlier coins. Hydroshacks represent a new wave of industrial scale technology that commenced with waterproof electronics and ended up creating a data center within shipping container walls.

Due to this fact efficiency improvements became a hotspot in such context like this one addressing the environmental impact of Bitcoin mining. By examining the technology, the miners have always been searching for ways to reduce power consumption while increasing computing power. Upgrading mining hardware is one way to achieve this goal by moving to more efficient chips known as ASICs (Application-Specific Integrated Circuits), that use less electricity for more computations. Doing this will therefore result in enhanced efficiency; decreased environmental impacts on Bitcoin mining activities.

## CHALLENGES AND LIMITATIONS

To achieve this, a few definitions and characterizations have been presented as a step towards more sustainable and efficient Bitcoin mining. The objective was not to define an ideal that is impossible for existing miners to attain, nor should it allow the bar to be set too low so that they would continue running at a cost without regard for preservation of the environment. However, transitioning towards sustainable practices in mining could be quite challenging. Some of these obstacles require an urgent response through policy for any long-term change to take place while others can be tackled later. For example, some of the challenges involved in the green movement when looking at Bitcoin mining have been identified. Among such are some issues concerning initial capital for a new renewable energy source and using second-hand greener hardware; not replacing current hardware or infrastructure with more environmentally friendly alternatives; one being tied down to specific hardware vendors; a huge maze consisting of different rules set by various national governments and tribes on earth. On one hand, blockchain use acts positively as it tends to oversee transparently and save ethical or responsible mining. However, it does not mean or do anything other than exist in such a way when information is being recorded. It does not create change, but reliance on the accuracy of information provided. Lastly, those in power or at the top of a hierarchy must ensure transparency by providing certainty about their plans and declarations. The switch to more sustainable mining practices may involve substantial investment in time, money, and strategic initiatives. Large scale miners are currently worse off because expenses are much higher than earnings; hence, the stakeholders' doubt has validity. So then, financiers are withholding funds for these miners to restructure, which leads to a vicious circle being drawn. Mining should always be approached with caution because of its unpredictable nature. Without major infrastructures or enough capital inflows, this could hardly happen since changes in hardware as well as mining practices may be almost impossible without the latter. On the other hand, other investors with the resources must put pressure on research institutions and governments to promote a movement towards change. It is recommended that the financial support goes to those companies which either manufacture retrofittable devices or design chips specifically for these purposes. Research with government policies concerning the mining sector in any given economy needs more attention than ever. Despite the enormity and complexities of these hurdles, the continued discussions on mining will address both possibilities for environmental responsibility and financial gains.

## THE FUTURE OF SUSTAINABLE BITCOIN MINING

Bitcoin mining has recently been coming into increased public attention with several regulatory proposals on how it can be controlled for its impact on carbon emissions. This has left many people wondering why no such thing has been done. This could mean that if they are ever forced to take responsibility for their carbon footprints, people will have to contend with many developments that may eventually create a completely sustainable Bitcoin network. Both directions aimed at reducing carbon emissions from Bitcoin can be termed innovative or regulatory.

Innovation refers to any technological changes that could make the eco-print size of Bitcoin miners significantly smaller. These could either be implemented voluntarily to preempt regulation or occur because of regulatory fines based on Bitcoin's use of carbon. Such changes are also themselves innovative immediately due to miners' intrinsic tendency to evade taxation. Bitcoin's fundamental inventions are already as green as typical banks and digital gold. It is possible to solve the elevation and cost problems posed by scaling with technology deployment options that would make Bitcoin mining the cleanest business ever on earth. Novel ASIC machines are being created which will help improve the energy efficiency of mining hardware & low-power consensus algorithms that don't damage the environment much.

The attraction of these technologies is not purely technical but also economic. Regulatory interventions, on the other hand, would increase the real cost of spinning ASICs. This might drive out dishonest miners who act in bad faith or push forward migration process when moving their operations to more enabled-regulatory environment jurisdictions around the world as long as Bitcoin networks function securely within their prescribed limits, technological advance will likely remain preferable over others. In terms of scalability, governance has already been established for how scaling is done within bitcoin Taking off from that point, it became clear that if demand supporting transactions became saturated, then new chains like Bitcoin Cash had to arise out there forthwith. A similar pattern could also be the case in the future should there arise any need to address environmental issues concerning Bitcoin mining itself. Gaining a wider acceptance among members public regarding its importance would imply some sort of sentiment gap between markets and miners as far as dealing with this problem is concerned.

The mining sector is at a critical stage in its history. The more concentrated the sector becomes in terms of stakeholders involved, the more decision-making power these people wield. Miners are facing perhaps the most critical phase of cryptocurrency since they grapple with the challenge of transitioning towards a more sustainable future. Calls for regulation or reduction of carbon use by Bitcoin are probable and could draw resistance from markets with ESG-conscious entities. In this context, it is possible that other than merely complying, the mining industry has an opportunity to take an active role based on the psychology behind ESG. As a result, companies would themselves assume leadership roles in advancing creative technologies that are also energy-efficient and environment-friendly. Through innovation alone can embodied carbon be reduced while increasing market cap, thus making Bitcoin mining among the greenest on earth.

## Technological Innovations

Technological innovations have increased the possibility of environmentally friendly mining of bitcoins.

### *Energy Efficient ASIC Miners*

ASIC miners, short for Application Specific Integrated Circuit miners, are generally thousands of times faster than traditional general-purpose hardware because they can perform only one task. The most popular manufacturer of these chips says its Antminer S19 Pro can process up to 110 trillion hashes per second while consuming 3250W from the power grid. Other smaller entities like Argo Technologies in London have also been increasing their use of software to optimize performance on their hardware in a bid to stay ahead of peers using methods such as air-cooling oxidation or immersion cooling. Among their submissions to the Bitcoin Mining Council, one of the highlighted institutions also mentioned about effective cooling solutions they use standard air-cooled data center facilities but are in the process of retrofitting a 210-megawatt facility with a cooling system designed to work in sub-zero temperatures that harvests the heat which is produced by hydroelectricity.

### *Alternative Consensus Mechanisms and Applications of Distributed Ledger Technology*

Adopting alternative consensus mechanisms could be one way forward; these might entail using ‘Proof of Stake’ or other less energy-consuming alternatives like ‘Proof of Authority,’ ‘Proof of Elapsed Time,’ or ‘Proof of History’ among others, instead of ‘Proof of Work’, which is associated with high energy consumption and requires long blockchain validation time. The adaptation could also be achieved by refining the Bitcoin code base concurrently with such measures as smart contract/side chains in Bitcoin Gold, adapting hash rate and transaction volume limits, full-blown tablet armies dedicated to examining off-chain solutions such as the Lightning Network, L2 Decentralized/L2 L.P., “Taproot,” Schnorr signatures, RE-BLND-combining newly developed concepts such as Taproot, MPP and Scallion together with other methods. The capacity exists for applying these technologies to testing ‘off-chain’ solutions involving ‘off-chain’ transactions such as lightning transactions and optimizing ‘off-chain’ solutions in general. The suggested solutions may involve studying faster block distribution methods, changing the rules for calculating mineral rewards, changing Gödel hash rates for converting some part of transactions fully away from miners, transaction minification and simpler validation in quicker blocks created at higher rates.

**TABLE 2**
**ECONOMIC IMPACT OF BITCOIN MINING IN RELATION TO SUSTAINABILITY**

| Year | Bitcoin Price (USD) | Bitcoins Mined (Millions) | Energy Consumption (TWh) | Carbon Emissions (Million Metric Tons $CO_2$) | E-Waste (Metric Tons) |
|---|---|---|---|---|---|
| 2014 | 500 | 4.0 | 3.0 | 255 | 500 |
| 2015 | 300 | 5.0 | 3.5 | 270 | 600 |
| 2016 | 450 | 5.5 | 4.0 | 279 | 700 |
| 2017 | 1000 | 7.0 | 5.0 | 304 | 900 |
| 2018 | 700 | 8.0 | 5.5 | 275 | 1,100 |
| 2019 | 900 | 9.0 | 6.0 | 275 | 1,300 |
| 2020 | 8,000 | 10.0 | 7.0 | 255 | 1,500 |
| 2021 | 50,000 | 12.0 | 8.5 | 270 | 1,800 |
| 2022 | 40,000 | 14.0 | 10.0 | 279 | 2,200 |
| 2023 | 30,000 | 16.0 | 11.5 | 304 | 2,700 |
| 2024 | 35,000 (Projected) | 18.0 (Projected) | 13.0 (Projected) | 275 (Projected) | 3,200 (Projected) |

*Source:* ***Bitcoin Price & Mined Data:*** *Blockchain.com;* ***Energy & Emissions:*** *CBECI; Jana et al (2022);* ***E-Waste:*** *Research studies*

When Bitcoin prices go up, more miners direct their funds into hardware and resources necessary for mining this popular cryptocurrency. This is well elaborated in figure 5 where it has been shown that how much power is consumed during the bitcoin mining process is highly dependent on the prices of bitcoin.

**FIGURE 5**
**BITCOIN PRICE AND CARBON EMISSIONS**

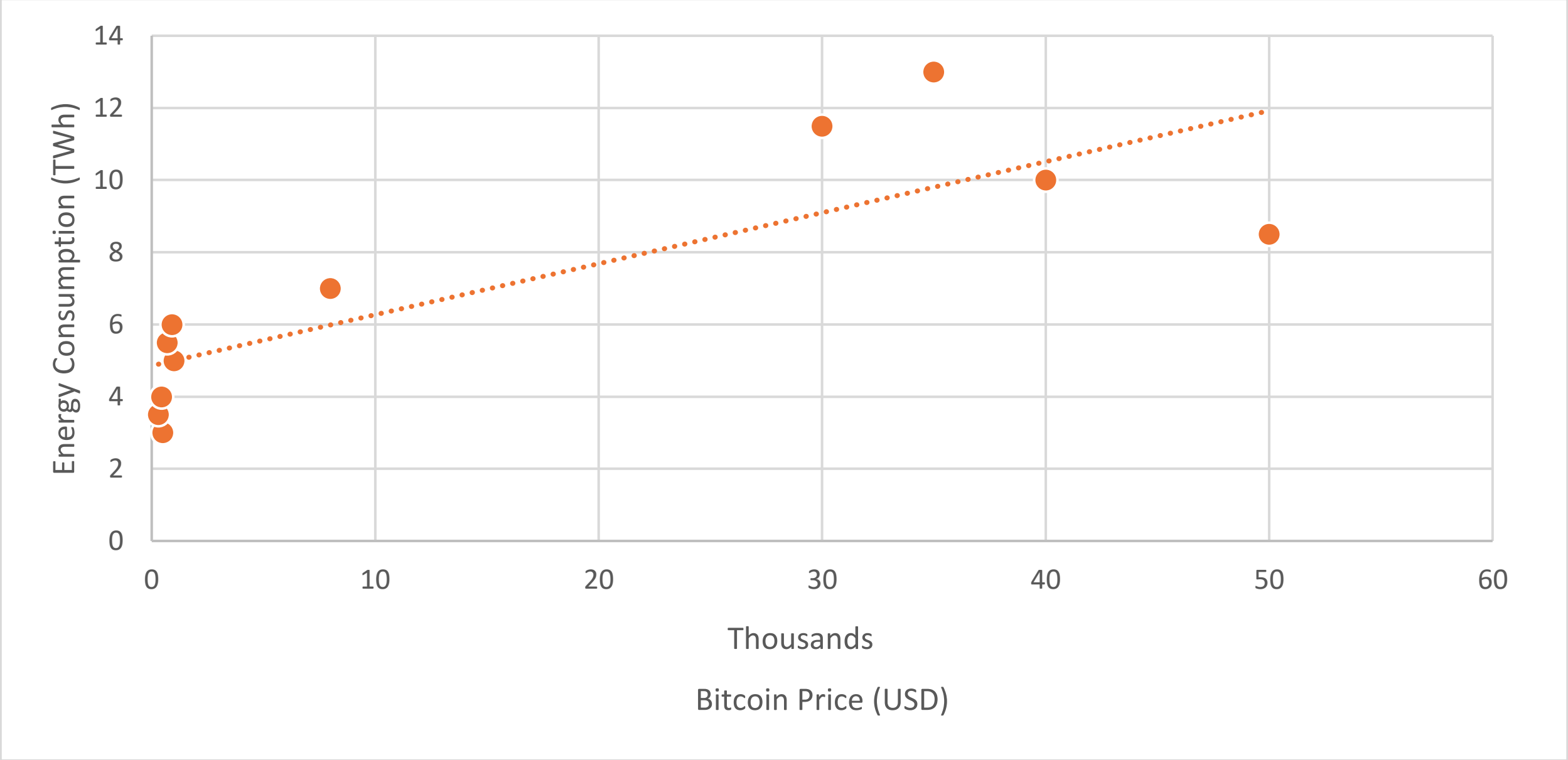

## Legal and Regulatory Changes

While bitcoins are decentralized, the local authority is planning intervention to make mining more sustainable. Despite the many shortcomings, one conspicuous advantage of having a regulatory framework is that it is possible for anyone to check the compliance of any miner. Some regulatory requirements may include stating or reducing carbon emissions. These will reduce costs of power for compliant mining operations through discounted electricity rates for miners who use or produce green power during mining.

Under the conditions set forth above, governments that impose emissions restrictions on goods and services indirectly govern both the amounts of carbon emission resulting from mining activities and computer energy consumed during mining work by making them correlate with the efficiency of computation required. Various costs and benefits that may be associated with adherence by miners to the emission standards have been highlighted while the likely punishment for defaulters was discussed too at length. Yet another key benefit of governance in place is its potential shield for operators who are aligned with the same law. Determining what is sustainable can be tricky but for those who are taking part and trying, it is all about finding out carbon emission at each unit of Bitcoin produced. It would be easy to come up with legislation for determining sustainability when it comes to mining process related to Bitcoins, this would be very beneficial for those miners who are already working hard to be sustainable.

## CONCLUSION

It is necessary to balance between innovativeness and sustainability in the context of Bitcoin mining for reduction from environmental impact caused while at the same time promoting technological growth. Bitcoin mining process consumes vast amounts of energy due in large part to its proof-of-work "mechanism" leading to significantly high levels of emissions and generation of electronic waste, however this sector has accepted such issues and now there are moves towards utilizing clean power sources integrating power efficient equipment for mining plus looking into different consensus mechanisms. Furthermore, there is hope that legislative frameworks together with other motivational strategies aimed at enhancing environmentally friendly practices during mining operations may serve as guides. To guarantee sustainability in Bitcoin mining so that our future generations can benefit from it without negative environmental implications, we require continuous policy initiatives, adoption of better technologies, and maintenance environmental standards adherence (html). These actions will not only mitigate immediate concerns about e-waste but also set us off on a course leading to creating a more resilient digital economy which can stand shocks if they occur. If this transformation is embraced, mining Bitcoins could be construed as an act of environmental responsibility without necessarily losing its position as an innovative activity economically and technologically. This model approach could be emulated in sectors where innovation intersects with sustainability, thus creating a more responsive technological environment.

## REFERENCES

Akbar, N.A., Muneer, A., ElHakim, N., & Fati, S.M. (2021). Distributed hybrid double-spending attack prevention mechanism for proof-of-work and proof-of-stake blockchain consensuses. *Future Internet*. Retrieved from mdpi.com.

Alfieri, F., Spiliotopoulos, C., & Takoudis, G. (2023). *ICT Task Force study*. Retrieved from europa.eu

Alharby, M. (2023). A dynamic block reward approach to improve the performance of blockchain systems. *PeerJ Computer Science*. Retrieved from peerj.com

Arora, P., & Nagpal, R. (2022, May). Blockchain technology and its applications: A systematic review of the literature. In *Proceedings of the International Conference on Innovative Computing & Communication* (ICICC). [HTML]

Awan, U., Sroufe, R., & Shahbaz, M. (2021). Industry 4.0 and the circular economy: A literature review and recommendations for future research. *Business Strategy and the Environment, 30*(4), 2038–2060. researchgate.net

Azman, M., & Sharma, K. (2020, August). HCH DEX: A secure cryptocurrency e-wallet & exchange system with two-way authentication. In *2020 Third International Conference on Smart Systems and Inventive Technology* (ICSSIT) (pp. 305–310). IEEE. [HTML]

Bachér, J., Rintala, L., & Horttanainen, M. (2022). The effect of crusher type on printed circuit board assemblies' liberation and dust generation from waste mobile phones. *Minerals Engineering*. Retrieved from sciencedirect.com

Badea, L., & Mungiu-Pupăzan, M.C. (2021). *The economic and environmental impact of bitcoin*. IEEE access. Retrieved from ieee.org

Bajra, U.Q., Rogova, E., & Avdiaj, S. (2024). Cryptocurrency blockchain and its carbon footprint: Anticipating future challenges. *Technology in Society*. [HTML]

Bastian-Pinto, C.L., Araujo, F.V.D.S., Brandão, L.E., & Gomes, L.L. (2021). Hedging renewable energy investments with Bitcoin mining. *Renewable and Sustainable Energy Reviews*, *138*, 110520. [HTML]

Bikubanya, D.L., & Radley, B. (2022). Productivity and profitability: Investigating the economic impact of gold mining mechanisation in Kamituga, DR Congo. *The Extractive Industries and Society*. Retrieved from sciencedirect.com

Bogdanov, D., Ram, M., Aghahosseini, A., Gulagi, A., Oyewo, A.S., Child, M., . . . Breyer, C. (2021). Low-cost renewable electricity as the key driver of the global energy transition towards sustainability. *Energy*, *227*, 120467. Retrieved from sciencedirect.com

Bukhari, M.Y., Ansari, A.A., Yousif, M., Hassan, M., & Hassan, U. (2024). Current and future implications of bitcoin mining on energy and climate change. *MRS Energy & Sustainability*, 1–14. [HTML]

Cao, C., Cen, Z., Feng, X., Wang, Z., & Zhu, Y. (2022). Straightforward guess and determine analysis based on genetic algorithm. *Journal of Systems Science and Complexity*, *35*(5), 1988–2003. [HTML]

Chaurasia, Y., Subramanian, V., & Gujar, S. (2021, December). PUPoW: A framework for designing blockchains with practically-useful-proof-of-work & vanitycoin. In *2021 IEEE International Conference on Blockchain* (Blockchain) (pp. 122–129). IEEE. [PDF]

Chen, M., & Ogunseitan, O.A. (2021). Zero E-waste: Regulatory impediments and blockchain imperatives. *Frontiers of Environmental Science & Engineering*, *15*, 1–10. Retrieved from escholarship.org

Cooke, S.L., & Xu, Y. (2024). *Trillion-dollar Stake Of Carbon Neutrality, The: Energy Infrastructures In Hong Kong And The Greater Bay Area*. Retrieved from worldscientific.com

Corbet, S., & Yarovaya, L. (2020). The environmental effects of cryptocurrencies. *Cryptocurrency and blockchain technolog*y. Retrieved from dcu.ie

Desislavov, R., Martínez-Plumed, F., & Hernández-Orallo, J. (2021). *Compute and energy consumption trends in deep learning inference*. arXiv preprint arXiv:2109.05472. [PDF]

Donovan, K.P. (2023). Decarbonizing Decentralized Currency: How Decentralized Datacenter Overlay Zones Could Ensure Bitcoin Mining's Clean Transition to the United States. *The Urban Lawyer*. [HTML]

Douaud, G., Lee, S., Alfaro-Almagro, F., Arthofer, C., Wang, C., McCarthy, P., ... & Smith, S.M. (2022). SARS-CoV-2 is associated with changes in brain structure in UK Biobank. *Nature*, *604*(7907), 697–707. Retrieved from nature.com

Erdiaw-Kwasie, M.O., Abunyewah, M., & Baah, C. (2024). A systematic review of the factors–Barriers, drivers, and technologies–Affecting e-waste urban mining: On the circular economy future of developing countries. *Journal of Cleaner Production*. Retrieved from cdu.edu.au

Erdin, E., Cebe, M., Akkaya, K., Bulut, E., & Uluagac, S. (2021). A scalable private Bitcoin payment channel network with privacy guarantees. *Journal of Network and Computer Applications*, *180*, 103021. Retrieved from sciencedirect.com

Farjana, S.H., Mungombe, T.M., Gamage, H.M.K., Rajwani, A.S., Tokede, O., & Ashraf, M. (2023). Circulating the E-Waste Recovery from the Construction and Demolition Industries: A Review. *Sustainability*, *15*(16), 12435. Retrieved from mdpi.com

Ferdous, M.S., Chowdhury, M.J.M., & Hoque, M.A. (2021). A survey of consensus algorithms in public blockchain systems for crypto-currencies. *Journal of Network and Computer Applications, 182*, 103035. Retrieved from latrobe.edu.au

Finney, B.R. (2024). *Win-Win Environmental Regulations for Crypto Mining: Developing a Regulatory Program That Reduces Environmental Harm and Promotes Innovation and …. BCL Re*v. Retrieved from bc.edu

Fryer, J., & Garcia, P. (2023). The Good, the Bad and the Ugly: Practices and Perspectives on Hardware Acceleration for Embedded Image Processing. *Journal of Signal Processing Systems*. [HTML]

Gallersdörfer, U., Klaaßen, L., & Stoll, C. (2020). Energy consumption of cryptocurrencies beyond bitcoin. *Joule*. Retrieved from cell.com

Ghosh, A., Gupta, S., Dua, A., & Kumar, N. (2020). Security of Cryptocurrencies in blockchain technology: State-of-art, challenges and future prospects. *Journal of Network and Computer Applications*, *163*, 102635. Retrieved from e-tarjome.com

Ghulam, S.T., & Abushammala, H. (2023). Challenges and opportunities in the management of electronic waste and its impact on human health and environment. *Sustainability*. Retrieved from mdpi.com

Goel, A., Masurkar, S., & Pathade, G.R. (2024). *An Overview of Digital Transformation and Environmental Sustainability: Threats, Opportunities and Solutions*. Retrieved from preprints.org

Gola, C., & Sedlmeir, J. (2022). Addressing the sustainability of distributed ledger technology. *Bank of Italy Occasional Paper*. Retrieved from uni.lu

Gunay, S., Sraieb, M.M., Kaskaloglu, K., & Yıldız, M.E. (2023). Cryptocurrencies and global sustainability: Do blockchained sectors have distinctive effects? *Journal of Cleaner Production, 425*, 138943. [HTML]

Gurgu, E., Zorzoliu, R.I., Pistol, L., Gurgu, I., Ungureanu, C., & Nae, G. (2021). The relationship between big data driven technologies and performance management strategies applied to companies in the hospitality, tourism & travel industry. Annals of Spiru Haret University. *Economic Series*, *21*(4), 97–136. [HTML]

Hallinan, K.P., Hao, L., Mulford, R., Bower, L., Russell, K., Mitchell, A., & Schroeder, A. (2023). Review and demonstration of the potential of bitcoin mining as a productive use of energy (PUE) to aid equitable investment in solar micro-and mini-grids worldwide. *Energies*, *16*(3), 1200. Retrieved from mdpi.com

Harooni, M.U. (2023). *The Relationship Between Bitcoin Price and Macroeconomic Variables*. Retrieved from researchgate.net

He, R., Li, M., Gan, V.J.L., & Ma, J. (2021). BIM-enabled computerized design and digital fabrication of industrialized buildings: A case study. *Journal of Cleaner Production*. Retrieved from google.com

Heinonen, H.T., Semenov, A., Veijalainen, J., & Hämäläinen, T. (2022). A survey on technologies which make bitcoin greener or more justified. *IEEE Access*, *10*, 74792–74814. Retrieved from ieee.org

Huang, H., Kong, W., Zhou, S., Zheng, Z., & Guo, S. (2021). A survey of state-of-the-art on blockchains: Theories, modelings, and tools. *ACM Computing Surveys (CSUR)*, *54*(2), 1–42. Retrieved from acm.org

Islam, M.R., Rashid, M.M., Rahman, M.A., & Mohamad, M.H.S.B. (2022). A comprehensive analysis of blockchain-based cryptocurrency mining impact on energy consumption. *International Journal of Advanced Computer Science and Applications*, *13*(4). Retrieved from researchgate.net

Jabłczyńska, M., Kosc, K., Ryś, P., Sakowski, P., Ślepaczuk, R., & Zakrzewski, G. (2023). Energy and cost efficiency of Bitcoin mining endeavor. *PloS one*, *18*(3), e0283687. Retrieved from plos.org

Jana, R.K., Ghosh, I., & Wallin, M.W. (2022). Taming energy and electronic waste generation in bitcoin mining: Insights from Facebook prophet and deep neural network. *Technological Forecasting and Social Change*, *178*, 121584. [HTML]

Jana, R.K., Ghosh, I., Das, D., & Dutta, A. (2021). Determinants of electronic waste generation in Bitcoin network: Evidence from the machine learning approach. *Technological Forecasting and Social Change*, *173*, 121101. Retrieved from uwasa.fi

Kayani, U., & Hasan, F. (2024). Unveiling Cryptocurrency Impact on Financial Markets and Traditional Banking Systems: Lessons for Sustainable Blockchain and Interdisciplinary Collaborations. *Journal of Risk and Financial Management*. Retrieved from mdpi.com

Khosravi, A., Sandoval, O.R., Taslimi, M.S., Sahrakorpi, T., Amorim, G., & Pabon, J.J.G. (2024). Review of energy efficiency and technological advancements in data center power systems. *Energy and Buildings*, 114834. [HTML]

Klassen, M., & Anthony, B.P. (2022). Legalization of Cannabis and agricultural frontier expansion. *Environmental Management*. [HTML]

Kumar, A., Sah, B.K., Mehrotra, T., & Rajput, G.K. (2023, April). A review on double spending problem in blockchain. In *2023 International Conference on Computational Intelligence and Sustainable Engineering Solutions* (CISES) (pp. 881–889). IEEE. [HTML]

Kuznetsova, E., & Anjos, M.F. (2020). Challenges in energy policies for the economic integration of prosumers in electric energy systems: A critical survey with a focus on Ontario (Canada). *Energy Poli*cy. [HTML]

Liu, Y., Chen, H., Zhang, L., Wu, X., & Wang, X. (2020). Energy consumption prediction and diagnosis of public buildings based on support vector machine learning: A case study in China. *Journal of Cleaner Production*. [HTML]

Lu, Y., Khan, Z.A., Alvarez-Alvarado, M.S., Zhang, Y., Huang, Z., & Imran, M. (2020). A critical review of sustainable energy policies for the promotion of renewable energy sources. *Sustainability*, *12*(12), 5078. Retrieved from mdpi.com

Luderer, G., Madeddu, S., Merfort, L., Ueckerdt, F., Pehl, M., Pietzcker, R., . . . Kriegler, E. (2022). Impact of declining renewable energy costs on electrification in low-emission scenarios. *Nature Energy*, *7*(1), 32–42. [HTML]

Lyu, Q., Qi, Y., Zhang, X., Liu, H., Wang, Q., & Zheng, N. (2020). SBAC: A secure blockchain-based access control framework for information-centric networking. *Journal of Network and Computer Applications*, *149*, 102444. [HTML]

Madaan, L., Kumar, A., & Bhushan, B. (2020, April). Working principle, application areas and challenges for blockchain technology. In *2020 IEEE 9th international conference on communication systems and network technologies* (CSNT) (pp. 254–259). IEEE. [HTML]

Maleš, U., Ramljak, D., Krüger, T.J., Davidović, T., Ostojić, D., & Haridas, A. (2023). Controlling the difficulty of combinatorial optimization problems for fair proof-of-useful-work-based blockchain consensus protocol. *Symmetry*, *15*(1), 140. mdpi.com

Malfuzi, A., Mehr, A.S., Rosen, M.A., Alharthi, M., & Kurilova, A.A. (2020). Economic viability of bitcoin mining using a renewable-based SOFC power system to supply the electrical power demand. *Energy*. [HTML]

Martins, R.M., & Gresse Von Wangenheim, C. (2023). Findings on teaching machine learning in high school: A ten-year systematic literature review. *Informatics in Education*. Retrieved from vu.lt

McClean, A., & Pedersen, O.W. (2023). The role of regulation in geothermal energy in the UK. *Energy Policy*. Retrieved from sciencedirect.com

McNally, K.F., & Kolivand, H. (2024). Comparative Analysis of Bitcoin Mining Machines and Their Global Environmental Impact. *EAI Endorsed Transactions on Scalable Information Systems*, *11*. Retrieved from ljmu.ac.uk

Melnyk, L., Derykolenko, O., Mazin, Y., Matsenko, O., & Piven, V. (2020). Modern trends in the development of renewable energy: The experience of the EU and leading countries of the world. *Mechanism of an Economic Regulation*, *3*(89), 117–133. Retrieved from http://mer-journal.sumy.ua/index.php/journal/article/view/64

Mendoza Beltran, A., Cox, B., Mutel, C., van Vuuren, D.P., Font Vivanco, D., Deetman, S., . . . Tukker, A. (2020). When the background matters: using scenarios from integrated assessment models in prospective life cycle assessment. *Journal of Industrial Ecology*, *24*(1), 64–79. Retrieved from wiley.com

Milutinović, V., Azer, E.S., Yoshimoto, K., Klimeck, G., Djordjevic, M., Kotlar, M., . . . Ratkovic, I. (2021, June). The ultimate dataflow for ultimate supercomputers-on-a-chip, for scientific computing, geo physics, complex mathematics, and information processing. In *2021 10th Mediterranean Conference on Embedded Computing* (MECO) (pp. 1–6). IEEE. [PDF]

Náñez Alonso, S.L., Jorge-Vázquez, J., Echarte Fernández, M.Á., & Reier Forradellas, R.F. (2021). Cryptocurrency mining from an economic and environmental perspective. Analysis of the most and least sustainable countries. *Energies*, *14*(14), 4254. Retrieved from mdpi.com

Niaz, H., Shams, M.H., Liu, J.J., & You, F. (2022). Mining bitcoins with carbon capture and renewable energy for carbon neutrality across states in the USA. *Energy & Environmental Science*. Retrieved from rsc.org

Okorie, D.I., Gnatchiglo, J.M., & Wesseh, P.K. (2024). Electricity and cryptocurrency mining: An empirical contribution. *Heliyon*. Retrieved from cell.com

Palma, J.G. (2023). *A fading "extractivist" model and more productivity-enhancing alternatives that just can't generate enough credibility—while populism looks for magical solutions*. Retrieved from cam.ac.uk

Peserico, N., Ferreira de Lima, T., Prucnal, P., & Sorger, V.J. (2022). Emerging devices and packaging strategies for electronic-photonic AI accelerators: Opinion. *Optical Materials Express*, *12*(4), 1347–1351. Retrieved from optica.org

Pouresmaieli, M., Ataei, M., Qarahasanlou, A.N., & Barabadi, A. (2023). Integration of renewable energy and sustainable development with strategic planning in the mining industry. *Results in Engineering, 20*, 101412. Retrieved from sciencedirect.com

Qi, R., Li, S., Qu, L., Sun, L., & Gong, C. (2020). Critical factors to green mining construction in China: A two-step fuzzy DEMATEL analysis of state-owned coal mining enterprises. *Journal of Cleaner Production*. [HTML]

Rahman, M.M., Khan, I., Field, D.L., Techato, K., & Alameh, K. (2022). Powering agriculture: Present status, future potential, and challenges of renewable energy applications. *Renewable Energy*. [HTML]

Razmjoo, A., Kaigutha, L.G., Rad, M.V., Marzband, M., Davarpanah, A., & Denai, M.J.R.E. (2021). A Technical analysis investigating energy sustainability utilizing reliable renewable energy sources to reduce CO2 emissions in a high potential area. *Renewable Energy*, *164*, 46–57. Retrieved from northumbria.ac.uk

Sandberg, K., & Chamberlin, S. (2023). *Web3 and Sustainability*. Linux Foundation Research: San Francisco, CA, USA. Retrieved from linuxfoundation.org

Sarkodie, S.A., Ahmed, M.Y., & Leirvik, T. (2022). Trade volume affects bitcoin energy consumption and carbon footprint. *Finance Research Letters*. Retrieved from sciencedirect.com

Schinckus, C. (2021). Proof-of-work based blockchain technology and Anthropocene: An undermined situation? *Renewable and Sustainable Energy Reviews*. [HTML]

Siddique, I., Smith, E., & Siddique, A. (2023). Assessing the sustainability of bitcoin mining: Comparative review of renewable energy sources. *Journal of Alternative and Renewable Energy Sources*, *10*(1), 10–46610. Retrieved from academia.edu

Singh, R., Tanwar, S., & Sharma, T.P. (2020). Utilization of blockchain for mitigating the distributed denial of service attacks. *Security and Privacy*. [HTML]

Sovacool, B.K., Upham, P., & Monyei, C.G. (2022). The "whole systems" energy sustainability of digitalization: Humanizing the community risks and benefits of Nordic datacenter development. *Energy Research & Social Science*. Retrieved from sciencedirect.com

Spurr, A., & Ausloos, M. (2021). Challenging practical features of Bitcoin by the main altcoins. *Quality & Quantity*. Retrieved from springer.com

Stoddard, I., Anderson, K., Capstick, S., Carton, W., Depledge, J., Facer, K., . . . Williams, M. (2021). Three decades of climate mitigation: why haven't we bent the global emissions curve? *Annual Review of Environment and Resources*, *46*(1), 653–689. Retrieved from annualreviews.org

Suárez, Á. (2023). *The Fundamentals of Bitcoin: The guide that will teach you the philosophical, economic, and technical fundamentals behind Bitcoin*. [HTML]

Suleyman, M. (2023). *The coming wave: Technology, power, and the twenty-first century's greatest dilemma*. Retrieved from amazonaws.com

Toorajipour, R., Oghazi, P., Sohrabpour, V., Patel, P.C., & Mostaghel, R. (2022). Block by block: A blockchain-based peer-to-peer business transaction for international trade. *Technological Forecasting and Social Change*, *180*, 121714. Retrieved from sciencedirect.com

Tsang, K.P., & Yang, Z. (2021). The market for bitcoin transactions. *Journal of International Financial Markets, Institutions and Money*, *71*, 101282. [HTML]

van der Giesen, C., Cucurachi, S., Guinée, J., Kramer, G.J., & Tukker, A. (2020). A critical view on the current application of LCA for new technologies and recommendations for improved practice. *Journal of Cleaner Production*, *259*, 120904. Retrieved from sciencedirect.com

Voumik, L.C., Islam, M.A., Ray, S., Mohamed Yusop, N.Y., & Ridzuan, A.R. (2023). CO2 emissions from renewable and non-renewable electricity generation sources in the G7 countries: Static and dynamic panel assessment. *Energies*, *16*(3), 1044. Retrieved from mdpi.com

Wang, Q., Guo, J., Li, R., & Jiang, X. (2023). Exploring the role of nuclear energy in the energy transition: A comparative perspective of the effects of coal, oil, natural gas, renewable energy, and nuclear power on …. *Environmental research*. [HTML]

Xiao, Z., Cui, S., Xiang, L., Liu, P.J., & Zhang, H. (2023). The environmental cost of cryptocurrency: Assessing carbon emissions from bitcoin mining in China. *Journal of Digital Economy*. Retrieved from sciencedirect.com

Yazıcı, A.F., Olcay, A.B., & Olcay, G.A. (2023). A framework for maintaining sustainable energy use in Bitcoin mining through switching efficient mining hardware. *Technological Forecasting and Social Change*, *190*, 122406. [HTML]